\documentclass[prl, amsmath, twocolumn, amssymb, aps, superscriptaddress, floatfix, preprintnumbers, notitlepage, 10pt]{revtex4-1}
\usepackage{times}
\usepackage{graphicx}
\usepackage{amsmath, braket, amsfonts}
\usepackage{amssymb}
\usepackage{natbib}
\usepackage{todonotes}
 \usepackage[modulo]{lineno}
\usepackage{bm, color, ulem}

\def\lsim{\lower -0.3ex \hbox{$<$} \kern -0.75em \lower 0.7ex \hbox{$\sim$}}
\def\gsim{\lower -0.3ex \hbox{$>$} \kern -0.75em \lower 0.7ex \hbox{$\sim$}}

\begin{document}

\title{Emergent Dirac gullies and gully-symmetry breaking quantum Hall states in ABA trilayer graphene}
\author{A.A. Zibrov}
\affiliation{Department of Physics, University of California, Santa Barbara CA 93106 USA}
\author{P. Rao}
\affiliation{Institute of Science and Technology, Am Campus 1, 3400 Klosterneuburg, Austria}
\author{C. Kometter}
\affiliation{Department of Physics, University of California, Santa Barbara CA 93106 USA}
\author{E. M. Spanton}
\affiliation{California Nanosystems Institute, University of California, Santa Barbara, CA 93106, USA}
\author{J.I.A. Li}
\affiliation{Department of Physics, Columbia University, New York NY 10025 USA}
\author{Cory R. Dean}
\affiliation{Department of Physics, Columbia University, New York NY 10025 USA}
\author{T. Taniguchi}
\affiliation{Advanced Materials Laboratory, National Institute for Materials Science, Tsukuba, Ibaraki 305-0044, Japan}
\author{K. Watanabe}
\affiliation{Advanced Materials Laboratory, National Institute for Materials Science, Tsukuba, Ibaraki 305-0044, Japan}
\author{M. Serbyn}
\affiliation{Institute of Science and Technology, Am Campus 1, 3400 Klosterneuburg, Austria}
	\author{A.F. Young}
\affiliation{Department of Physics, University of California, Santa Barbara CA 93106 USA}

\begin{abstract}
We report on quantum capacitance measurements of high quality, graphite- and hexagonal boron nitride encapsulated Bernal stacked trilayer graphene devices. At zero applied magnetic field, we observe a number of electron density- and electrical displacement-tuned features in the electronic compressibility associated with changes in Fermi surface topology.
At high displacement field and low density, strong trigonal warping gives rise to three new emergent Dirac cones in each valley, which we term  `gullies.' The gullies are centered around the corners of hexagonal  Brillouin zone and related by three-fold rotation symmetry.  At low magnetic fields of $B=1.25$~T, the gullies manifest as a change in the degeneracy of the Landau levels from two to three. Weak incompressible states are also observed at integer filling within these triplets Landau levels, which a Hartree-Fock analysis indicates are associated with Coulomb-driven nematic phases that spontaneously break  rotation symmetry.
\end{abstract}

\maketitle

In graphene multilayers, strong trigonal warping of the electronic band structure leads to a complex evolution of Fermi surface topology within the low energy valleys located at the corners of the hexagonal Brillouin zone\cite{mccann_landau-level_2006,shi_tunable_2018}.
The comparatively small energy scales characterizing the underlying interlayer hopping processes ($\sim100$ meV) renders these transitions accessible via electrostatic gating, providing a highly tunable platform for engineering both zero- and high magnetic field electronic structure. Of particular interest is the possibility to use band structure engineering to create novel manifolds of degenerate Landau levels (LLs), where enhanced electron-electron interaction effects can lead to novel correlated ground states.
However, such control comes at the cost of requiring high sample quality to avoid smearing the subtle electronic features.

\begin{figure*}
\begin{center}
\includegraphics[width=6.5 in]{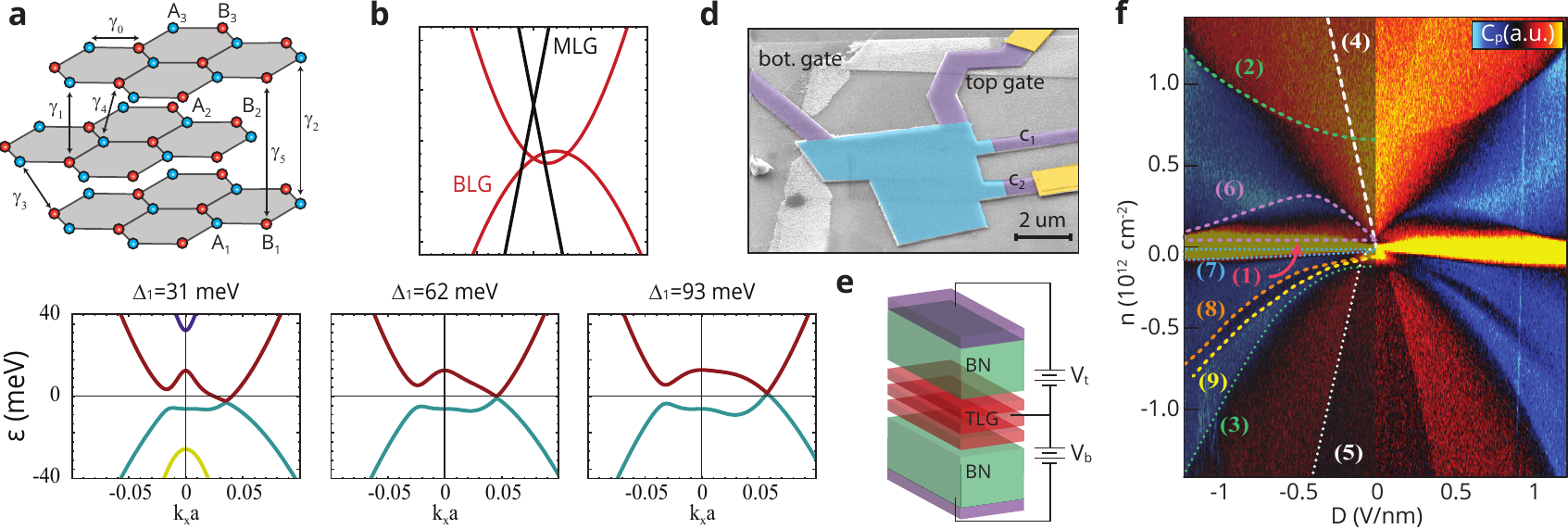}
\caption{
\textbf{Trilayer graphene band structure and penetration field capacitance measurements at $B=0$.}
\textbf{a.} Lattice structure of ABA trilayer graphene with hopping parameters identified. In addition to the $\gamma_i$, the electronic structure is determined by the interlayer potentials $\Delta_1\propto D$ and the relative potential of the inner layer with respect to the outer layers, $\Delta_2$.
\textbf{b.} Electronic band structure of trilayer graphene in the absence of an applied displacement field.  The linear monolayer-like and parabolic bilayer-like bands are labeled. The momentum is relative to the $K$ point in the $\vec k_x\parallel \Gamma-K$ direction.
\textbf{c.} Band structure evolution under applied electric field.  For a wide range of electric fields, the low energy structure is described by three isolated Dirac cones slightly displaced from the $K$($K'$) points.
\textbf{d.} False color electron micrograph  of the measured trilayer graphene device.  The active region is indicated in cyan.
\textbf{e.} Device schematic: trilayer graphene encapsulated in $\sim 20$ nm BN with few-layer graphite top and bottom gates. Independent contacts to the gates and graphene layer allow independent control of charge density $n=c_t V_t+c_b V_b$ and displacement electric field $D=\epsilon_{hBN} (V_t/d_t-V_b/d_b)$, where $\epsilon_{hBN}\approx 3$ and $d_{t(b)}=18,20$ nm are the distances to the gates.
\textbf{f.} Penetration field capacitance $C_P$ at $B=0$~T and $T\approx 50$~mK as a function of $n$ and $D$. $D$ breaks the mirror symmetry of the ABA-stacked trilayer graphene and induces an on-site energy difference $\Delta_1$ between the outer layers.
Main features visible in the experimental data are indicated by dashed lines and numerals. The $D<0$ region is shaded to increase the visibility of the features. Data is plotted on a saturated color scale (see Fig. S5).
\label{b0}
}
\end{center}
\end{figure*}

In this Letter we report magnetocapacitance measurements of exceptionally high quality Bernal-stacked (ABA) trilayer graphene devices (Fig. \ref{b0}a).  Absent an applied perpendicular electric field, the band structure of ABA trilayer is described by independent monolayer graphene-like (linear) and bilayer graphene-like (parabolic) bands\cite{koshino_gate-induced_2009,partoens_graphene_2006, avetisyan_electric-field_2009} in each of the two valleys centered at the high symmetry $K$ and $K'$ points (Fig. \ref{b0}b). Applied electric displacement field $\vec D$ strongly hybridizes these two sectors, driving the linear monolayer-like band to high energies and generating new structure in the low-energy bilayer-like bands (Fig. \ref{b0}c).  For large electric fields, the strong trigonal warping is predicted to lead to the formation of new Dirac gullies centered around each of the two original valleys\cite{serbyn_new_2013,morimoto_gate-induced_2013} and are related to each other by three-fold rotation symmetry. At quantizing magnetic fields, the three-fold symmetry of the gullies may lead to novel broken symmetry ground states\cite{sodemann_quantum_2017}, including nematic states as recently observed on the surface of high purity bismuth crystals\cite{feldman_observation_2016}.

Past experiments on ABA trilayer graphene\cite{craciun_trilayer_2009,kumar_integer_2011,henriksen_quantum_2012,taychatanapat_quantum_2011,bao_stacking-dependent_2011,lee_broken_2013,campos_quantum_2012,shimazaki_landau_2016,stepanov_tunable_2016,datta_strong_2017}
have observed  features associated with numerous aspects of the single particle band structure, including a variety of electric- and magnetic-field tuned LL crossings\cite{koshino_landau_2011,yuan_landau_2011} that tightly constrain band structure parameters\cite{taychatanapat_quantum_2011,campos_quantum_2012}.  Recent experiments have also found evidence for interaction-induced quantum Hall ferromagnetic states at high magnetic field\cite{lee_broken_2013,stepanov_tunable_2016,datta_strong_2017}. However, the high-electric field regime of the  Dirac gullies has not been explored in high mobility devices where interaction driven states might be accessible.

To access the high mobility, high-$D$ regime, we study ABA trilayer flakes encapsulated in hexagonal boron nitride dielectric layers and single-crystal graphite gates\cite{zibrov_tunable_2017} (Fig. 1d). We use few-layer graphite to contact the trilayer, allowing us to vary both the total charge density and displacement field $\vec D$ across the trilayer (Fig. 1e). We measure the penetration field capacitance $C_P$\cite{eisenstein_negative_1992}, defined as the capacitance between top and bottom gate with the graphene layer held at constant potential.
The finite density of states $\partial n/\partial \mu$ of the trilayer partially screens the electric field between the top and bottom gate, reducing the measured $C_P$ so that (for top- and bottom gates with geometric capacitance $c$) $C_P=c^2/(2c+\partial n/\partial \mu)\propto\left(\partial n/\partial \mu\right)^{-1}$ for $\partial n/\partial \mu\gg c$. Changes in $C_P$ are thus associated with changes in the Fermi surface size or topology. 

Fig. \ref{b0}f shows $C_P$ measured at $B$=0 as a function of $D$ and electron density $n$. A variety of $n$ and $D$-tuned discontinuities are readily visible and indicated in the Figure with numeric labels (1)-(9).  These include a sharp $C_P$ maximum at charge neutrality for both positive and negative $D$ (1); two elevated $C_P$ features with parabolic boundaries  at negative and positive $n$ (2-3), two low-$C_P$ regions with triangular boundary within the parabolic regions (4-5), a `wing'-shaped high $C_P$ region both above and below charge neutrality (6-7), and a narrow elevated $C_P$ region that runs parallel to the parabolic feature for negative $n$ bounded by contours (8-9).
Some of the capacitance features can be associated with the single-particle band-structure by inspection. For example, (1) is consistent with the small band gap or linear band crossing expected at charge neutrality\cite{serbyn_new_2013}.  Features (4-5), meanwhile, are  identified as the  extrema of the linear bands (purple and yellow in Fig. \ref{b0}c) which disperse rapidly to high energy with increasing $D$.  Additional features are thus associated with the complex band minima of the low energy bands.

To understand the remaining observed compressibility features we perform tight binding simulations of the trilayer graphene band structure.  Energy eigenvalues are computed using a 6-band tight binding model (see Supplementary information).  Hopping between different atoms within the unit cell is parameterized by six tight binding parameters $\gamma_i,~i=1..6$, one on-site energy $\delta$, and two energy asymmetries $\Delta_1$ and $\Delta_2$.
$\Delta_1$ describes the potential difference between the top and bottom layers and is most directly tuned by the strength of an externally applied polarizing electric field $D$. $\Delta_2$ measures the potential imbalance between the central layer and the two outer layers, and screening effects within the trilayer.

\begin{figure*}
\begin{center}
\includegraphics[width=6.5 in]{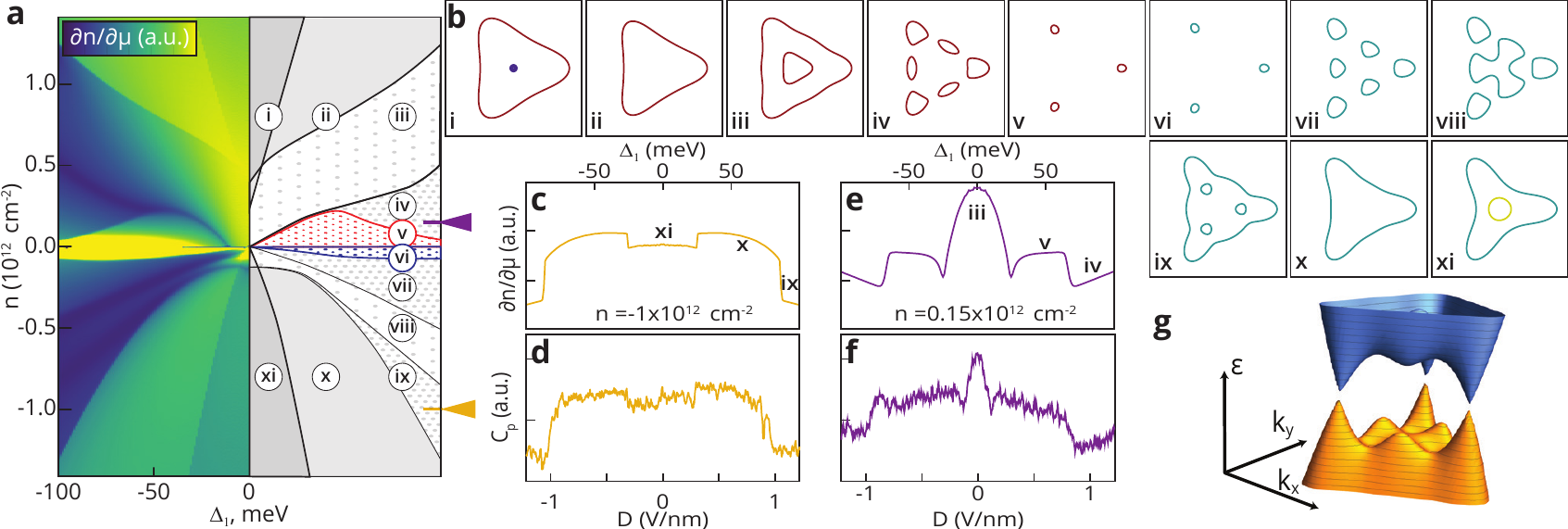}
\caption{
\textbf{
Fermi surface topology}
\textbf{a.} Left: Inverse electronic compressibility $\partial n/\partial \mu$ calculated from a 6-band tight binding Hamiltonian as a function of interlayer asymmetry $\Delta_1$ and electron density $n$. Right: schematic showing regions (indexed by the Roman numerals) separated by sharp changes in the compressibility.
\textbf{b.} Fermi contours calculated at each of the points indexed by roman numerals in a.  Color indicates the band and follows the convention of Fig. 1c; note that panels i-v are Fermi surfaces of electrons while vi-xi are Fermi surfaces of holes.
\textbf{c.} Simulated $\partial n/\partial \mu$ and
\textbf{d.} measured $C_P$ at $n=-1\times 10^{12} \text{cm}^{-2}$.  The discontinuous jump in the data at $D\approx\pm$.3~V/nm coincides with population of the 2nd hole subband (xi-x transition), while the jump at $D\approx\pm$ .95~V/nm coincides with the opening of internal electron-like Fermi surfaces within the main hole pocket (x-ix transition).
\textbf{e.} Simulated $\partial n/\partial \mu$ and
\textbf{f.} measured $C_P$ at $n=.15\times 10^{12} \text{cm}^{-2}$.  The sharp minimum at $D\approx\pm$.5 V/nm coincides with a Lifshitz transition from one multiply-connected electron pocket (iii) to three disconnected Dirac cones (v). At the discontinuity at $D\approx\pm$.9 V/nm, the Dirac cones are joined by three additional auxiliary pockets.
\textbf{g.} Band structure near $K$-point for $\Delta_1=75$ meV showing the emergent Dirac gullies.
\label{fermi}
}
\end{center}
\end{figure*}

Figure \ref{fermi}a shows the calculated inverse compressibility within this model, as a function of the carrier density and $\Delta_1\propto |\vec D|$.  Both the geometric and parasitic capacitances within  the device influence the mapping of $\partial n/\partial \mu\leftrightarrow C_P$ between calculated compressibility and measured data.  Moreover, interactions likely renormalize the compressibility particularly when it is high.  We thus restrict ourselves to qualitative comparisons of the magnitude of the signals, and plot both in arbitrary units.  We do, however, achieve quantitative agreement between data and simulation for the \textit{position} of extrema and discontinuities for parameters $\gamma_0=3.1$, $\gamma_1=.38$, $\gamma_2=-0.021(5)$, $\gamma_3=0.29$, $\gamma_4=0.141(40)$, $\gamma_5=0.050(5)$, $\delta=0.0355(45)$, and $\Delta_2=0.0035$, where all energies are expressed in eV.
Notably, the model succeeds in matching the experimentally observed features only for an exceptionally narrow range of parameters,
providing tighter constraints on $\{\gamma_i\}$ and $\{\Delta_i\}$ than previously achieved using only LL coincidences\cite{taychatanapat_quantum_2011,campos_landau_2016,shimazaki_landau_2016}. In addition to the parameters $\gamma_i$ and $\Delta_2$, a single scale factor $\alpha=.165~\mathrm{e}\cdot$nm is chosen so that $\Delta_1=\alpha \cdot D$.  $\alpha$ describes dielectric screening of the perpendicular electric field by the trilayer, implying an effective $\epsilon^\perp_{TLG}\approx 4$ for the trilayer itself (see supplementary information).

The agreement between theory and experiment allows us to understand the connection between the observed compressibility features and the nature of the Fermi contours.  Fig.~\ref{fermi}b shows calculated Fermi surface contours in 11 distinct regions throughout the experimentally accessed parameter regime.  Regions (i) and (xi), for example, are distinguished by the existence of a second, independent Fermi surface arising from the second electron- or hole-subband, respectively, as intuited above.  All other regions are separated by Lifshitz transitions and distinguished by differences in Fermi surface topology within a single electron- or hole-band. We note that signatures of Lifshitz transitions were recently found in tetralayer graphene\cite{shi_tunable_2018} at zero magnetic field, but  no direct compressibility measurements of Lifshitz transitions have been reported.
With the exception of regions iii-iv, all of the regions are bounded by experimentally observed features described in Fig. 1.  We note that features characterized by a diverging density of states, such as the iii-iv boundary, only weakly modify the measured capacitance and are barely discernible even in Fig. \ref{b0}f.

Fig. 2c-d shows comparisons of traces from the measured capacitance and the numerically calculated inverse compressibility at $n=-1.0\times 10^{12} \text{cm}^{-2}$. Both data and simulation show matching discontinuities associated with the band edge of the second hole subband (i.e., the xi-x transition) as well as the nucleation of three new electron pockets within the main hole-like Fermi pocket (x-ix).
Of particular interest is the regime of  low $n$ and large $D$, where the gully Dirac points are predicted\cite{serbyn_new_2013}.  Fig. \ref{fermi}e-f show line traces at $n=.15\times 10^{12} \text{cm}^{-2}$.  The `wing' region,  bounded by sharp discontinuities in both the measured signal and simulated data, is readily identified with region (v), in which the Fermi surface arises from three isolated gully Dirac cones (Fig. \ref{fermi}g).

\begin{figure*}
\begin{center}
\includegraphics[width=6.5 in]{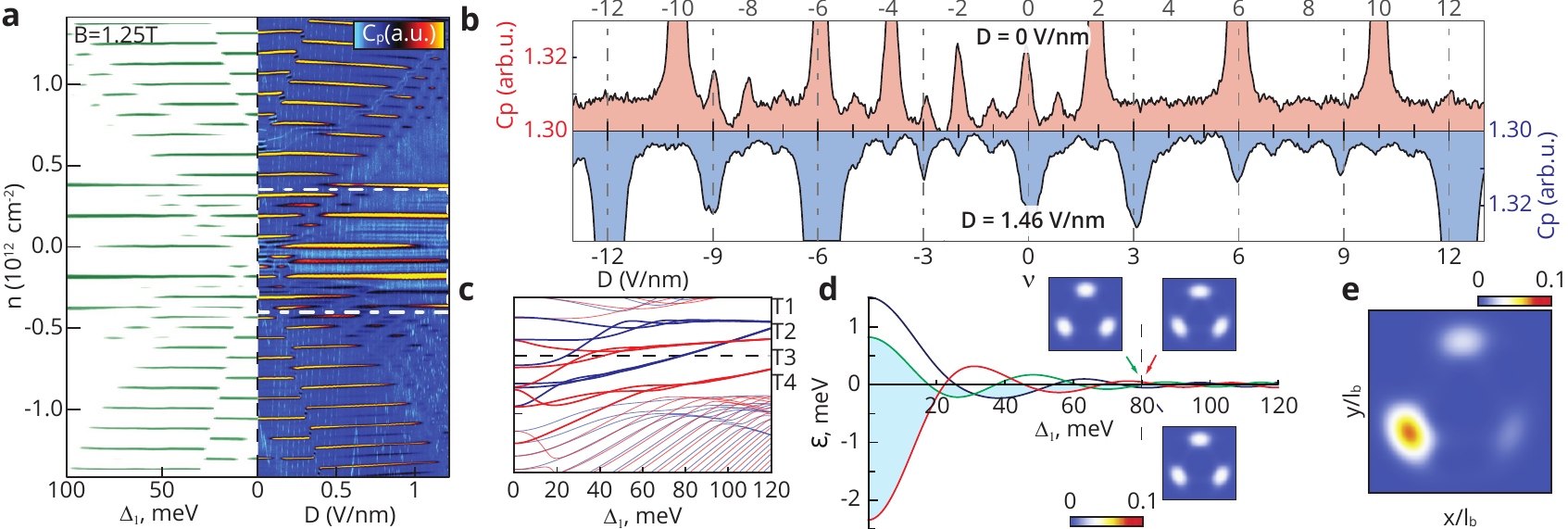}
\caption{
\textbf{Triplet Landau levels and interaction induced symmetry breaking.}
\textbf{a.} Right panel: Penetration field capacitance $C_P$ measured at $B$=1.25 T as a function of $D$ and $n$. The dashed lines indicate the region of low carrier density near the valence and conduction band minima where  trigonal warping has strongest effect and leads to a formation of new Dirac points.  Left panel: simulated inverse compressibility at $B$=1.25 T based on band structure parameters. A phenomenological thermal broadening of 0.1meV is assumed to generate contrast, so that only the largest gaps are visible in green.
\textbf{b.} $C_P$ traces for $n \in [-0.5,0.5]\times 10^{12}$ cm$^{-2}$ at $D=0$ (red) and at $D=1.46$ V/nm (blue). The $D=0$ line trace shows strong capacitance peaks at even filling factors, in contrast to the peaks at multiples of three ($\nu=\pm 3, 6, 9, 12$) for $D=1.46$ V/nm.
\textbf{c.} Evolution of LLs  at $B=1.25 $ T as a function of interlayer potential difference $\Delta_1$.  As the electric potential increases, 12 distinct LLs at $\Delta_1\approx0$ intertwine into 4 quasi-degenerate triplets, denoted $T_1\dots T_4$, separated from a near- continuum of closely spaces LLs by energy gaps.
\textbf{d.} Expanded view of the triplet T2 with the average energy of the triplet subtracted.  Insets show the real-space probability distribution for a coherent state formed from wave-functions in each of the component LLs. All respect rotation symmetry.
\textbf{e.} Real-space probability distribution of the Hartree-Fock ground state at 1/3 filling of the spin-polarized triplet T2 at $\Delta_1=80$~meV, showing strongly broken three-fold rotation symmetry.
\label{triplets} }
\end{center}
\end{figure*}

In addition to its thermodynamic signatures at $B=0$, the emergence of isolated Dirac cones can be expected to lead to new transport, optical, and thermodynamic phenomenology at finite magnetic fields.  In monolayer graphene, for example, the two inequivalent valleys lead to four-fold internal degeneracy of the  LLs, with an additional factor of two arising from electron spin. The observation of four-fold degeneracy was a critical feature of the first experimental demonstrations of the Dirac spectrum in monolayer graphene\cite{zhang_experimental_2005,novoselov_two-dimensional_2005}.

The gully Dirac cones similarly manifest as increased LL degeneracy.  Figure 3a shows $C_P$ data measured at B=1.25~T alongside the results of diagonalizing the trilayer Hamiltonian in the presence of a magnetic field (simulations ignore spin splitting; see supplementary information).  Larger energy gaps manifest as prominent peaks in $C_P$ at filling factors $\nu=eBn/h$, spaced by integer multiples of $g$, the internal LL degeneracy. Near $D=0$, we observe the strongest capacitance peaks spaced by $\Delta\nu=2$, in agreement with the two-fold valley degeneracy ($g=2$) but lifted spin degeneracy (Fig.~\ref{triplets}b, top). In contrast  at large displacement fields ($D>0.7$ V/nm) and near charge neutrality---i.e., in the regime of the Dirac gullies---this behavior changes, with the most prominent gaps spaced by $\Delta\nu=3$ for $-12<\nu<12$ (see Fig.~\ref{triplets}b, bottom). The calculated single particle energy spectrum (Fig~\ref{triplets}c) shows that displacement field leads to the formation of four triplets of LLs per spin projection (labeled T1, T2, T3, and T4); within each triplet, three LLs intertwine into a single three-fold quasi-degenerate band consistent with the observed LL degeneracy. We note that triplet LLs are a generic feature of trigonally warped multilayer band structures, and evidence for three-fold degenerate LLs has previously been reported in suspended bilayer graphene samples\cite{varlet_anomalous_2014}.

While the observation of triplet LLs is consistent with expectations from our single-particle model, close examination of high $D$ data reveals departures from the noninteracting picture.  In particular, we observe $C_P$ peaks at all integer filling factors $-6<\nu<12$, corresponding to the dashed region of Fig. \ref{triplets}a (see also Fig. S6), including weak peaks at $(\nu~\mathrm{mod}~3)\neq 0$.  These gaps persist \textit{without} closing over the whole range of $D>0.7$ V/nm.
This is qualitatively inconsistent with the single particle spectrum, which predicts that within each triplet (T1...T4 in Fig. \ref{triplets}c) the single particle eigenstates evolve via a series of crossings with increasing $\Delta_1$ (Fig. \ref{triplets}d).  One thus expects these anomalous gaps to undergo repeated closings, in contrast to their observed persistence.

The failure of the single-particle picture is not surprising.  The estimated bandwidth of each triplet (Fig. \ref{triplets}d), $\delta\varepsilon<0.5$ meV, is smaller than the scale of the Coulomb interactions, $E_C = e^2/(\epsilon \ell_B)\approx 10$ meV at $B=1.25$ T (here $e$ is the elementary charge, $\epsilon=6.6$ the in-plane dielectric constant of hBN\cite{geick_normal_1966}, and $\ell_B=\sqrt{\hbar/(eB)}$ the magnetic length). Taking these interactions into account, the individual LLs within the triplet are effectively degenerate; the ground state at integer filling must result from minimizing repulsive interactions and is likely to result in a gapped, symmetry breaking quantum Hall ferromagnetic state.

We investigate this quantitatively using a variational Hartree-Fock analysis (see supplementary information) of the ground state when only one out of 3 LLs within a single spin branch of triplet T2 is filled (1/3 filling).  The three insets to Fig.~\ref{triplets}d show real space probability distributions for coherent states constructed for each of the three components of T2. Absent interactions, the ground state at 1/3 filling consists of the lower energy component of T2 for a given value of $B$ and $\Delta_1$, and preserves rotation symmetry. In contrast, the Hartree-Fock ground state
(Fig.~\ref{triplets}e) spontaneously breaks the $C_3$ symmetry--it is a gully nematic. As long as $\delta\varepsilon\ll E_C$, the gap will be only weakly moduated by $\Delta_1$, making it insensitive to the single-particle level crossings, in agreement with experimental observation.

The nematic ground state is merely one example of a  symmetry breaking channel.
Intuitively, nematics are favored by interactions when LL wave functions are localized in well separated real-space pockets, as in the case in the highly anisotropic wave functions of Fig.~ \ref{triplets}e.  In a momentum space picture, these pockets are associated with the main Dirac gullies represented in the contours of Fig.~\ref{fermi}b v-vi.  In this limit, ABA trilayer triplet LLs resemble the case of the (111) surface of SnTe recently considered theoretically.~\cite{li_su3_2016}.
Our single-particle calculations suggest that other limiting behaviors can also be realized in ABA trilayer graphene, resulting in  qualitatively different ground states. For instance, the triplet states T1 and T4 are considerably less anisotropic, being associated with multiple momentum space pockets close to the $K$($K'$) points as in Fig~\ref{fermi}b vii. In these triplets, isotropic ground states constructed from a superposition of triplet wavefunctions may be favored.  Notably, the relevant anisotropies within each triplet are continuously tunable by external electric and magnetic fields, making ABA trilayer graphene an remarkably versatile platform for exploring correlation effects in unusual quantum Hall ferromagnets.  Cataloging the theoretical possibilities, and determining how to distinguish them experimentally, will be the topic of future work.

\bibliographystyle{unsrt}

\clearpage
\newcommand{\be}{\begin{equation}}
\newcommand{\ee}{\end{equation}}
\newcommand{\bea}{\begin{eqnarray}}
\newcommand{\eea}{\end{eqnarray}}
\newcommand{\HH}{{\cal H}}
\newcommand{\variphi}{\phi}
\newcommand{\LL}{{\cal L}}
\newcommand{\KK}{{\cal K}}
\newcommand{\VV}{{\cal V}}
\newcommand{\GG}{{\sf G}}
\newcommand{\tr}{{\rm tr\/}\,}
\newcommand{\p}{\partial}
\newcommand{\s}{\sigma}
\newcommand{\la}{\langle}
\newcommand{\ra}{\rangle}
\newcommand{\lb}{\left[}
\newcommand{\rb}{\right]}
\newcommand{\lp}{\left(}
\newcommand{\rp}{\right)}

\newcommand{\E}{{\cal E}}
\newcommand{\der}{{\partial}}
\renewcommand{\phi}{\varphi}
\renewcommand{\epsilon}{\varepsilon}
\renewcommand{\vec}[1]{{\bf #1}}
\def\nn{\nonumber\\}
\newcommand{\mpar}[1]{\marginpar{\small \it #1}}
\newcommand{\corr}[1]{\langle{ #1}\rangle}
\newcommand{\corrt}[1]{\langle{ #1}\rangle_\text{typ}}
\newcommand{\corrr}[1]{\langle\langle{ #1}\rangle\rangle}
\newcommand{\ins}[1]{|{ #1}\rangle}
\newcommand{\out}[1]{\langle{ #1}|}
\newcommand{\kp}{{\kappa'}}
\newcommand{\kpp}{{\kappa}}

\renewcommand{\emph}{\textit}
\renewcommand{\thefigure}{S\arabic{figure}}
\renewcommand{\thetable}{S\arabic{table}}
\renewcommand{\theequation}{S\arabic{equation}}

\begin{widetext}
\vspace*{0.5cm}
\textbf{\large Supplementary Online Material: Emergent Dirac gullies and gully-symmetry breaking quantum Hall states in ABA trilayer graphene}
\vspace*{0.5cm}
\end{widetext}
\tableofcontents
\section*{}
\indent In this supplementary materials we summarize the tight-binding description of ABA-stacked trilayer graphene and outline the procedure used to simulate the density of states. In addition, we discuss how we constrain and refine the tight-binding parameters using zero field and Landau level data. Finally, we provide details on the self-consistent Hartree-Fock calculation of symmetry broken states in Landau levels  and discuss their visualization. Supplementary figures S5 and S6, referenced in the main text, can be found on page S8.
\setcounter{equation}{0}
	\setcounter{figure}{0}
	\setcounter{table}{0}
	\setcounter{page}{1}
	\setcounter{section}{0}
	\makeatletter
	\renewcommand{\theequation}{S\arabic{equation}}
	\renewcommand{\thefigure}{S\arabic{figure}}
	\renewcommand{\thesection}{S\Roman{section}}
	\renewcommand{\thepage}{S\arabic{page}}
	\renewcommand*{\bibnumfmt}[1]{[S#1]}
	\twocolumngrid
	\section{A. Model and Methods}
	
	\subsection{A.1 Hamiltonian and bandstructure}
	To describe the band structure of ABA trilayer graphene we use the Slonczewski-Weiss-McClure parametrization of the tight-binding model~\cite{dresselhaus_intercalation_2002}. This parametrization uses six tight-binding parameters $\gamma_0\ldots\gamma_5$ to describe hopping matrix elements between different atoms:
	\begin{subequations} \label{Eq:gammas-def}
		\bea
		A_i\leftrightarrow B_i : \gamma_0,\quad &\qquad&
		B_{1,3}\leftrightarrow A_{2} : \gamma_1,
		\\
		A_1\leftrightarrow A_3 : \frac12 \gamma_2,
		&\qquad&
		A_{1,3}\leftrightarrow B_2 :  \gamma_3,
		\\
		\begin{matrix}
			A_{1,3}\leftrightarrow A_2\\
			B_{1,3}\leftrightarrow B_2
		\end{matrix}:  -\gamma_4,
		&\qquad&
		\ \  B_1\leftrightarrow B_3 :  \frac12 \gamma_5,
		\eea
	\end{subequations}
	where $A_i$ ($B_i$) refers to an atom from $A$ ($B$) sublattice, and index $i=1\ldots 3$ labels three layers~(see Fig.~1 in the main text). In addition, parameter $\delta$ accounts for an extra on-site potential for $B_1$, $A_2$, and $B_3$ sites, which are on top of each other. Finally, we use two additional parameters $\Delta_{1,2}$ to capture the effect of external electric field and charge asymmetry between internal and external layers of ABA graphene. Parameters $\Delta_{1,2}$ are related~\cite{lu_influence_2006,guinea_electronic_2006,min_ab_2007,koshino_gate-induced_2009} to layer potentials  $U_{1}\ldots U_{3}$ as:
	\begin{equation}   \label{Eq:D1D2def}
	\Delta_1  =  (-e)\frac{U_1-U_2}{2},
	\qquad
	\Delta_2
	=
	(-e)\frac{U_1-2U_2+U_3}{6}.
	\end{equation}
	We note that the above parameterization is spin-independent. As we shall see below, spin-degenerate simulations fully capture experimental features at zero magnetic field, and adequately describe Landau level data except in vicinity of neutrality point. Effects that break spin degeneracy, i.e. Zeeman splitting and electron interactions, are included only in Section~C where we address symmetry broken states in Landau levels.
	
	Via rotation of basis, the tight-binding Hamiltonian for ABA-stacking trilayer graphene can be decoupled into monolayer-graphene-like (SLG) and bilayer-graphene-like (BLG) sectors which are coupled due to presence of displacement field $\Delta_1$:
	\begin{equation}\label{Hamiltonian-full}
	H = \begin{pmatrix}H_{SLG} & V_{\Delta_1} \\ V^T_{\Delta_1} & H_{BLG}\end{pmatrix},
	\end{equation}
	where the matrix blocks are defined as:
	\begin{widetext}
\begin{eqnarray}\label{Hamiltonian}
	H_\text{SLG}&= &
	\begin{pmatrix} \Delta_2 - \frac{\gamma_2}{2} & v_0 \pi^{\dagger} \\
	v_0 \pi & -\frac{\gamma_5}{2}+\delta + \Delta_2 \end{pmatrix}, \\ \label{HamBLG}
	H_\text{BLG} &= &\begin{pmatrix}
	\frac{\gamma_2}{2}+ \Delta_2 & \sqrt{2} v_3 \pi & -\sqrt{2} v_4 \pi^{\dagger} & v_0 \pi^{\dagger}\\ \sqrt{2} v_3 \pi^{\dagger} & -2\Delta_2 &  v_0 \pi & -\sqrt{2} v_4 \pi \\
	-\sqrt{2} v_4 \pi & v_0 \pi^{\dagger} & \delta-2\Delta_2& \sqrt{2} \gamma_1 \\
	v_0\pi & -\sqrt{2} v_4 \pi^{\dagger}& \sqrt{2}\gamma_1 & \frac{\gamma_5}{2} + \delta + \Delta_2
	\end{pmatrix},\\
	V_{\Delta_1} & =& \begin{pmatrix}
	\Delta_1&0&0&0 \\ 0&0&0&\Delta_1
	\end{pmatrix}.
	\end{eqnarray}
	\end{widetext}
	Here we introduced velocities $v_i= \sqrt{3}a \gamma_i/(2\hbar) $ corresponding to some of the tight-binding hopping matrix elements, where $a=0.246$~nm is the lattice constant. These notations coincide with those used in Ref.~\cite{serbyn_new_2013}. At zero magnetic field $B$, the operator $\pi$ in Eqs.~(\ref{Hamiltonian})-(\ref{HamBLG}) can be written as $\pi= \xi k_x + i k_y$, where $k$ is crystal momentum measured with respect to corresponding $K^\pm$ point labeled by $\xi=\pm1$. For finite magnetic field, $\pi$ can be replaced with the annihilation (creation) operator acting in the basis of Landau level states in the $K^-$ ($K^+$) valley.
	
	The capacitance measurements presented in this paper are sensitive to the band structure within a range $\sim 10$~meV from neutrality point.  Within this energy range one can obtain  additional insights into effects of TB parameters by deriving $2\times2$ low energy effective Hamiltonian of $H_\text{BLG}$. This Hamiltonian is obtained by projecting out 2 bands which are split by energies of order $0.5$~eV  away from neutrality point:
	\begin{equation}\label{Heff}
	\begin{aligned}
	H^\text{eff}_\text{BLG} = -\frac{1}{2m}\begin{pmatrix}0 & \pi^{\dagger 2}\\ \pi ^2 & 0 \end{pmatrix} +\sqrt{2} v_3\begin{pmatrix} 0 & \pi \\ \pi^\dagger& 0 \end{pmatrix} + \\ + \begin{pmatrix} \frac{\gamma_2}{2} + \Delta_2 & 0 \\ 0 & -2\Delta_2 \end{pmatrix} + ...
	\end{aligned}
	\end{equation}
	where $1/(2m) = v^2_0 / (\sqrt{2}\gamma_1)  [1 + O ({\gamma_4}/{\gamma_0})^2]$. We see that, to first order, $\gamma_4$ doesn't appear in the effective Hamiltonian and its effect on the band structure is small.
	
	From explicit form of $2\times2$ Hamiltonians for monolayer and bilayer blocks, Eqs.~(\ref{Hamiltonian}) and (\ref{Heff}) one can qualitatively understand the effects of tight-binding parameters on the band structure. The nearest neighbor hopping~$\gamma_0$ gives the fermi-velocity of the massless monolayer sector fermions. Interlayer hopping $\gamma_1$ influences to the effective mass of the bilayer graphene.  The trigonal warping term $\gamma_3$ determines the behavior of bilayer bands at small momenta.  Finally, small parameters $\Delta_2$, $\gamma_2$, $\delta$ and $\gamma_5$ located on the diagonal of Hamiltonians~(\ref{Hamiltonian}) and (\ref{Heff}) determine the magnitude of band gap and relative displacement of BLG and SLG bands.
	
	\subsection{A2. Simulation method}
	
	At zero magnetic field, we numerically calculate the charge density $n(\mu)$ and density of states~(DOS) $\nu(\mu)$ as a function of the external potential $\Delta_1$ and chemical potential $\mu$. We discretize the crystal momentum in vicinity of a given $K$ point. For a fixed value of $\Delta_1$ we calculate single particle energies for each point of the momentum grid by numerically diagonalizing the Hamiltonian~(\ref{Hamiltonian-full}). Density  $n(\mu)$~(density of states $\nu(\mu)$) is calculated by summing the Fermi-distribution $n_{F}(\varepsilon_{\bm k} - \mu)$~(derivative of Fermi function $n'_{F}(\varepsilon_{\bm k} - \mu)$) over all points in the grid,
	\begin{eqnarray}\label{n}
	n(\mu) &=& 4 g_\text{sym} \frac{S_k} {(2\pi)^2}  \frac{1}{N} \sum_{\bm k} n_F(\varepsilon_{\bm k} - \mu),\\ \label{nu}
	\nu(\mu) &=& 4 g_\text{sym} \frac{S_k} {(2\pi)^2}  \frac{1}{N} \sum_{\bm k} n'_F(\varepsilon_{\bm k} - \mu),
	\end{eqnarray}
	where $N = \sum_{\bm k} 1$ is the total number of momentum points in the considered portion of the Brillouin zone with area $S_k$, and factor of $4$ accommodates for spin and valley degeneracies. Finally, $g_\text{sym}$ takes into account the symmetry of the BZ: for example, $g_\text{sym}= 6$ for our simulations where we use the triangular grid covering 1/6 of vicinity of $K$ point. The normalization constant in Eqs.~(\ref{n})-(\ref{nu}) is chosen so that $n(\mu)$ and $\nu(\mu)$ have physical units m$^{-2}$ and m$^{-2}$ eV$^{-1}$ respectively.
	
	\begin{table*}[t]
		\begin{tabular}{ l | l | l | l | l | l | l | l | l }
			\hline\hline
			Data set & $\gamma_0$  & $\gamma_1$ & $\gamma_2$ & $\gamma_3$ & $\gamma_4$ & $\gamma_5$ & $\delta$ & $\Delta_2$ \\ \hline
			\cite{taychatanapat_quantum_2011} (Graphite) & 3.16 & 0.39 & -0.02 & 0.315 & 0.044 & 0.038& 0.037 & n/a \\
			\cite{gruneis_tight-binding_2008} (Graphite) & 3.0121 & 0.3077 & -0.0154 & 0.2583 & 0.1735& 0.0294&0.0214&n/a \\ 		
			\cite{shimazaki_landau_2016} & 3.1 & 0.39 &  -0.028 & 0.315 & 0.041 & 0.05 & 0.034& 0 \\
			\cite{campos_landau_2016} & 3.1& 0.39 & -0.02 to-0.016 & 0.315 & 0.04 to 0.14 &0.005 to 0.015&0.012 to 0.018 & n/a \\
			\cite{datta_strong_2017}& 3.1 & 0.39 & -0.028 & n/a &n/a& 0.01&0.021& n/a \\
			\cite{datta_landau_2018}& 3.1 & 0.39 & -0.02 & 0.315 & 0.12 & 0.018 & 0.02& 0.0043 to 0.0044 \\
			This paper & 3.1  & 0.38$\pm$0.003 & -0.021$\pm$0.005 & 0.29 & 0.141$\pm$0.04 & 0.05$\pm$0.005 & 0.0355$\pm$0.0045 & 0.0035 \\
			\hline
			\hline
		\end{tabular}
		\caption{\label{TB}Different sets of tight-binding parameters from the literature are compared to the set of parameters determined in this work. All parameters are given in units of eV,  ``n/a'' means that corresponding reference did not consider the corresponding parameter.}
	\end{table*}
	
	Simulations of DOS $\nu(\mu)$ in the presence of magnetic field $B=1.25$~T are carried out in two steps. First, we determine the Landau level spectrum $\varepsilon_n(\Delta_1)$ in each of the valleys, $K^+$ and $K^-$, as a function of displacement field. The LL spectrum is calculated via exact diagonalization of the Hamiltonian~(\ref{Hamiltonian-full}) with operators $\pi$, $\pi^\dagger$ replaced by properly truncated ladder operators~(see e.g.\ Ref.~\cite{serbyn_new_2013} for additional details).
	
	Next, we assume that each Landau level (LL) contributes a Lorentzian-shaped DOS centered at its energy. The total DOS is calculated as a sum of DOS from all LLs:
	\begin{eqnarray}
	\nu(\mu) &=& \sum_n \nu_n(\mu), \\ \label{nu-n}
	\nu_n(\mu) &=& 2\frac{eB}{2\pi\hbar c}  \frac{\Gamma}{[\mu-\varepsilon_n(\Delta_1)]^2 + \Gamma^2},
	\end{eqnarray}
	where factor ${eB}/({2\pi\hbar c})$ accounts for the LL degeneracy and $\Gamma$ is the LL broadening. Due to the small value of Zeeman splitting, we do not incorporate it in our calculation and treat all LLs as spin-degenerate. In order to account for this degeneracy, we include additional factor of 2 in Eq.~(\ref{nu-n}). Density $n(\mu)$ and density of states $\nu(\mu)$ are then calculated by summing individual contributions from all filled LLs for a grid in space of parameters $(\Delta_1,\mu)$. We used value of $\Gamma=0.1$~meV for our simulations.
	
	
	\section{B. Refinement of tight-binding parameters}
	
	The determination of tight-binding parameters for ABA trilayer graphene was performed by a number of earlier works~\cite{serbyn_new_2013,taychatanapat_quantum_2011,gruneis_tight-binding_2008,shimazaki_landau_2016,campos_landau_2016,datta_strong_2017,datta_landau_2018}. The resulting sets of tight-binding parameters are summarized in the Table~\ref{TB}. This table shows that despite overall consensus, values of some parameters differ quite significantly between different references.
	
	We use our zero field data and LL data to refine the existing parameter sets. We perform refinement of tight-binding parameters starting with values established in Ref.~\cite{serbyn_new_2013}. The tight-binding parameters are divided in two classes:
	\begin{enumerate}
		\item[(i)] Parameters $\gamma_0$, $\gamma_3$, $\gamma_4$ and $\Delta_2$ which influence $\nu(\mu)$ (measured via penetration field capacitance) at zero magnetic field.
		\item[(ii)] Parameters $\gamma_2$, $\delta$ and $\gamma_5$, which determine gaps in bilayer/monolayer sectors and thus can be constrained using Landau levels.
	\end{enumerate}
	After determining constraints from experimental data for the  Landau levels, we refine parameters in the set (i) using our simulations at zero magnetic field.
	
\begin{figure*}
		\includegraphics[width=\linewidth]{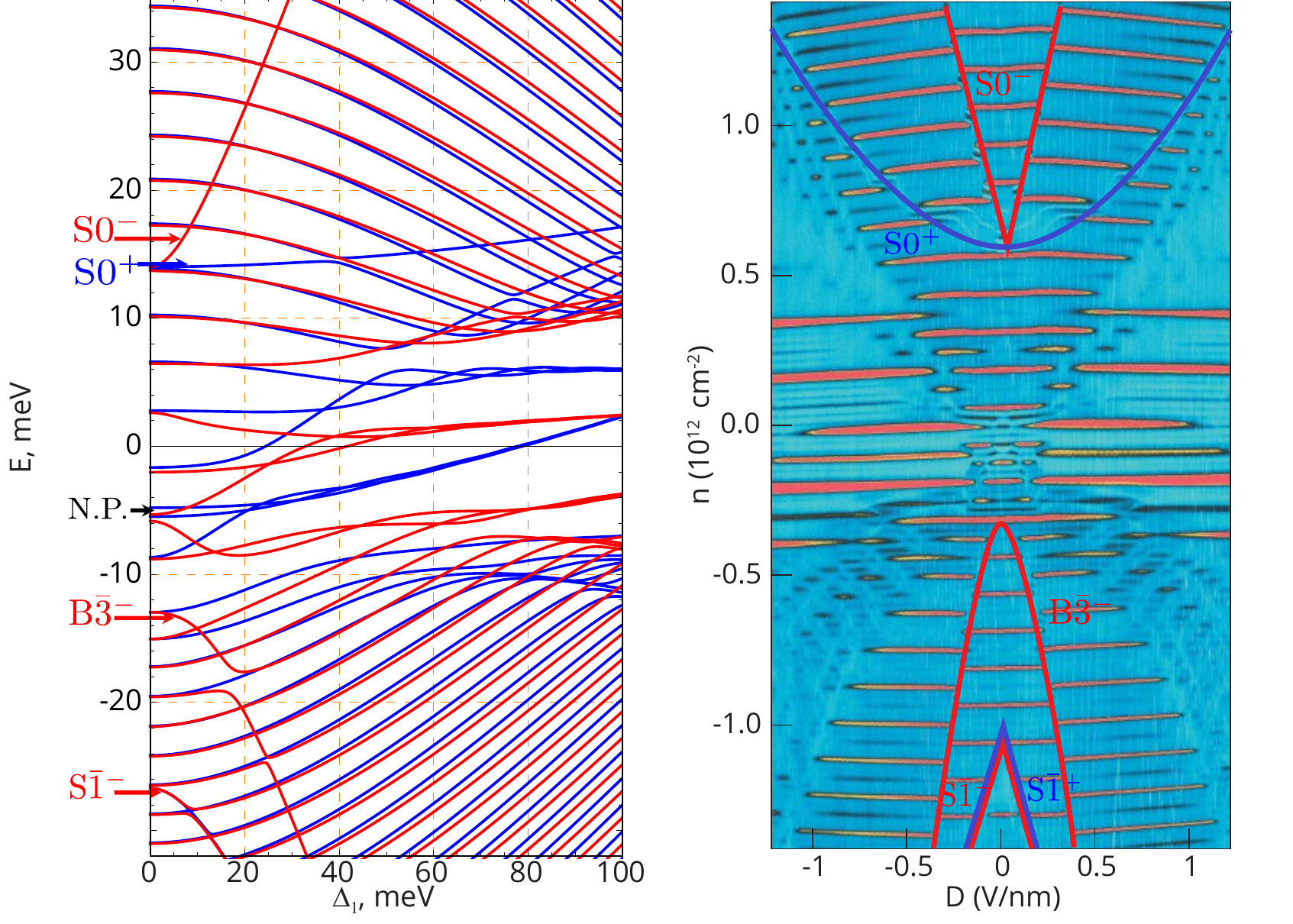}
		\caption{\label{bandstructureM} (left) LL fan diagram shows energies of LLs as a function of displacement field $\Delta_1$ at $B=1.25$~T. Blue (red) lines denote LLs from $K^+$ ($K^-$) valley. (right) Experimental data from the main text. The LL responsible for most prominent crossings are labeled explicitly.}
	\end{figure*}

	
	Magnetic field data imposes strict conditions on the tight-binding parameters $\gamma_2$, $\delta$ and $\gamma_5$. They must be chosen to satisfy the requirements that prominent LLs  have the correct positions corresponding to experimental data. Figure~\ref{bandstructureM} illustrates the positions of special LLs which are used to deduce the constraints on the tight-binding parameters. The LLs in Fig.~\ref{bandstructureM} are labeled as S$n^{\pm}$ or  B$n^{\pm}$ for $n=0,1$ and for $n\geq 2$ on the electron side, where letter specifies if the given LL belongs to SLG~(S) or BLG~(B) sector when the displacement field is vanishing, $\Delta_1=0$. We use bar above  LL indices to distinguish the LL on the hole-doped side. For example B$3^+$ (B$\bar 3^+$) stands for the LL with $n=3$ from bilayer sector on the electron (hole) doped side in $K^+$ valley. From comparing LL fan diagram to experimental data in Fig.~\ref{bandstructureM} we obtain the following requirements:
	\begin{enumerate}
		\item[(i)] At $\Delta_1=0$, there should be 9 (spin-degenerate) LLs between neutrality point (NP) and S$0^+$. Likewise, there are 5 (spin-degenerate) LLs between NP and B$\bar{3}^-$. In addition, LLs S$0^{\pm}$ and B$6^{\pm}$ are almost four-fold degenerate.
		\item[(ii)] Gap at neutrality point should vanish as a function of dispacement field $\Delta_1$. The most natural scenario for this is the touching of new emergent Dirac points, see Ref.~\cite{serbyn_new_2013}.
		\item[(iii)] LLs B$\bar{8}^{\pm}$ and S$\bar{1}^{\pm}$ are degenerate at $\Delta_1=0$; in addition there are 10 (spin degenerate) BLG LLs between B$\bar{3}^{\pm}$ and B$\bar{8}^{\pm}$.
	\end{enumerate}
	In order to use condition (i) we calculate the energies of relevant LLs. From Eqs.~(\ref{Hamiltonian}) and (\ref{Heff}) we find that the energies of $S0^{-}$ and $B0^{-}$ are given by $\pm \gamma_2/2 + \Delta_2$ respectively. Thus the number of LLs  between $S0^{-}$ and NP, which is close to $B0^{-}$, is controlled by parameter $\gamma_2$. To satisfy condition (i), this parameter should take the value $\gamma_2= 0.02\pm0.005$~eV which also results in the correct counting for $B\bar{3}^-$. In order to determine the associated error bars, we fix the value of all other parameters as their final values (see Table~\ref{TB}), determine the range of $\gamma_2$ where condition (1) is still satisfied. The values of $\gamma_2$ in the range $-0.016\le\gamma_2\le-0.025$~eV give the correct total 14 LLs between $B\bar{3}^-$ and $S0^-$. Thus, we determine
	\begin{equation}\label{Eq:g2sm}
	\gamma_2= -0.02\pm0.005~\text{eV}.
	\end{equation}
	
	Next, we determine parameter $\gamma_5$ from condition (ii) which implies the triplet crossing (see the main text). Increasing parameter $\gamma_5$ shifts the this crossing to smaller values of electric fields $\Delta_1$. In order to satisfy condition (2), we adjust
	\begin{equation}
	\gamma_5 = 0.05 \pm 0.005~\text{eV},
	\end{equation}
	where error bar is estimated by comparing the position of triplet crossing relative to crossings between LL S$0^+$ with B$12^\pm$ and B$11^\pm$.
	
	After we fix parameters $\gamma_{2,5}$,  $\delta$ must be chosen to satisfy the second part of condition (i). We see from Eq.~(\ref{Hamiltonian}) that energies of S$0^{\pm}$ LLs are $-\gamma_5/2 + \delta+\Delta_2$ and $\Delta_2-\gamma_2/2$ respectively. Thus, we obtain one condition which allows us to express $\delta$ via $\gamma_{2,5}$: $-\gamma_5/2 + \delta = -\gamma_2/2$. From here we determine
	\begin{equation}
	\delta = 0.0355\pm 0.0045~\text{eV},
	\end{equation}
	where we estimated error bars from known error bars for parameters~$\gamma_{2,5}$.
	
	Finally, to satisfy condition (iii), we need to adjust the parameter $\gamma_1$ by the small amount compared to its value in the literature. Decreasing $\gamma_1$ to be $\gamma_1=0.38$~eV increases the cyclotron frequency of the bilayer sector, resulting in the correct counts of LL number between  B$\bar{3}^{\pm}$ and S$\bar{1}^{\pm}$. By checking the range of $\gamma_1$ which gives correct crossing pattern between S$\bar{1}^{\pm}$ and B$\bar{8}^{\pm}$, and assuming LL width of $0.1$~meV, we determine the error bar as
	\begin{equation}
	\gamma_1=0.38\pm 0.003~\text{eV}.
	\end{equation}

\begin{figure}[h!]
		\includegraphics[width=\linewidth]{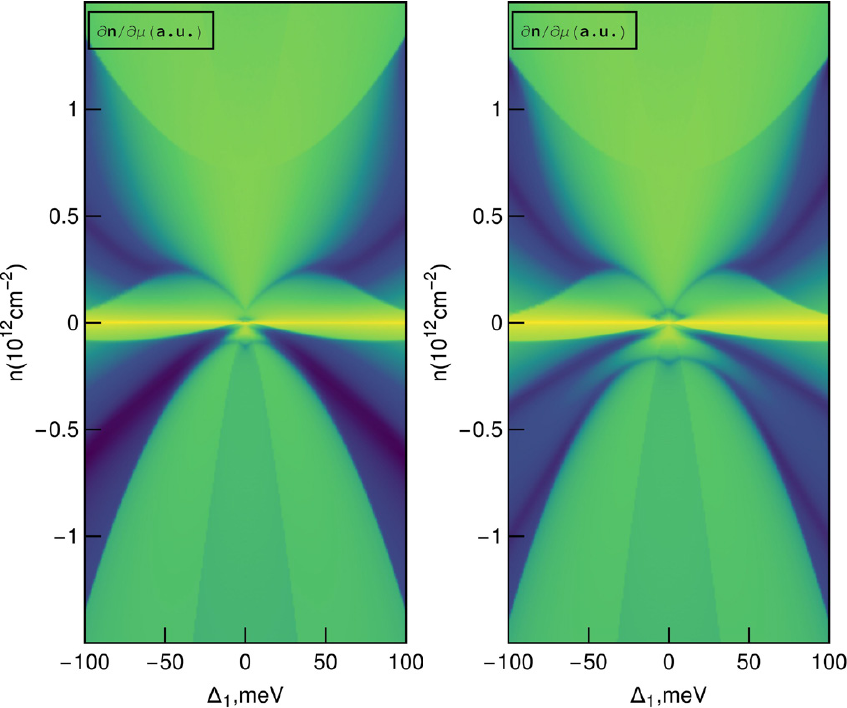}
		\caption{\label{g3} Increasing $\gamma_3$ from 0.25~eV (left) to 0.35~eV (right) increases the distance between the tip of feature (3) and the origin in the simulations, and also reduces the DOS near Lifshitz transitions at (8) and (9). Here the features are labeled in the same way as in the main text Fig.~1f.}
	\end{figure}

\begin{figure}
\begin{center}
		\includegraphics[width=\linewidth]{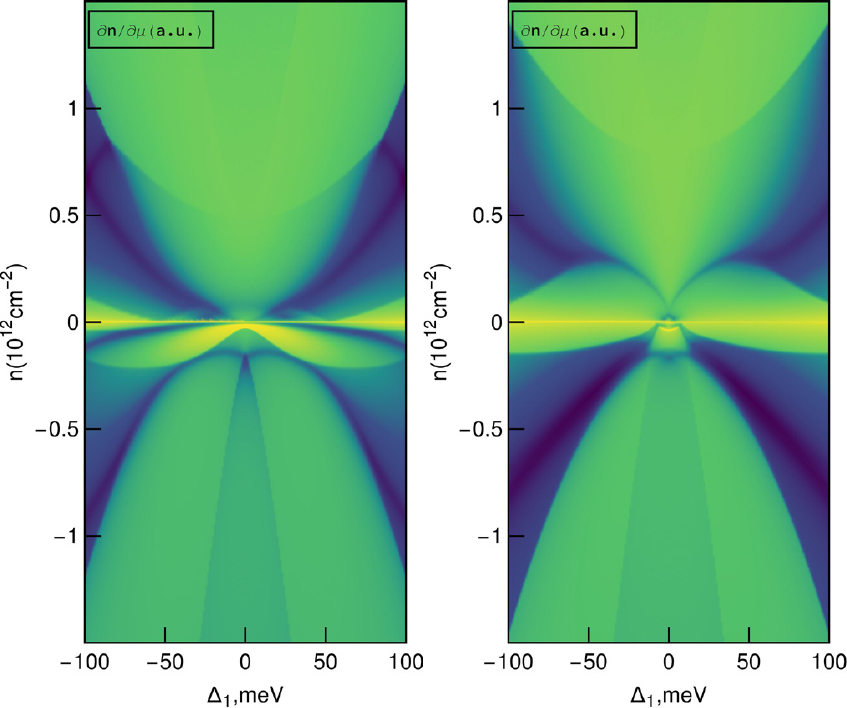}
		\caption{Increasing $\Delta_2$ from 0 (left) to 5 meV (right) pulls almost all features on the electron-doped side away from the neutrality point. In addition, upon increasing $\Delta_2$ the first Lifshits transition on the hole side (8) is displaced away from the neutrality point towards more negative fillings. \label{D2} }
\end{center}
	\end{figure}

\begin{figure}
		\includegraphics[width=\linewidth]{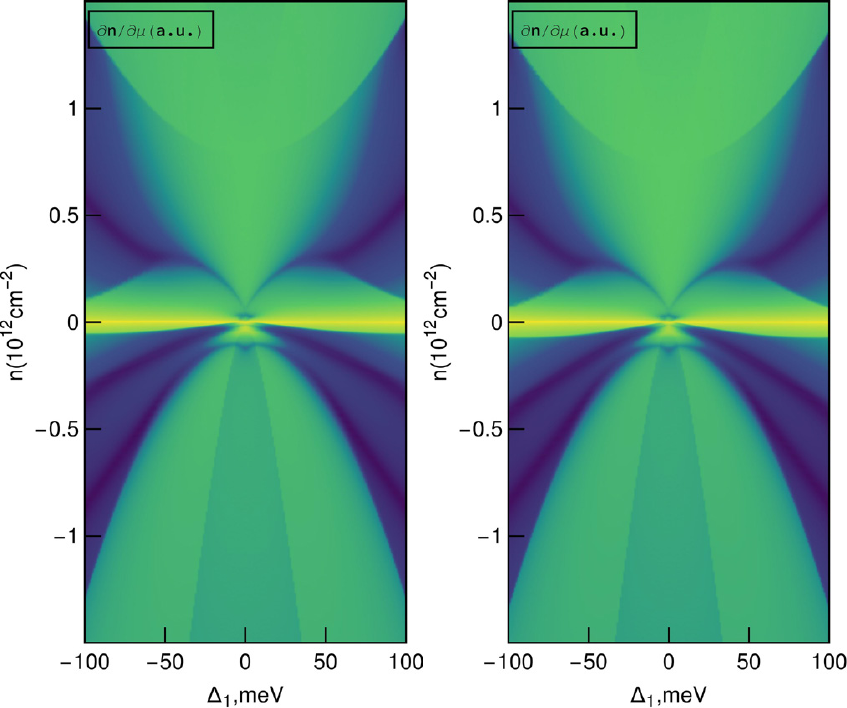}
		\caption{\label{g4} Upon increasing  $\gamma_4$ from $\gamma_4=0.041$~eV (left) to $0.1$~eV (right) Lifshits transitions (8) and (9) move closer to each other. }
	\end{figure}

	After determining parameters $\gamma_2$, $\gamma_5$,  $\delta$, and adjusting parameter $\gamma_1$ using LL data, we fix the remaining parameters $\gamma_3$, $\gamma_4$ and $\Delta_2$ by matching features in the DOS at zero magnetic field. Here we label the qualitative band features with the same notation as Fig.~1f in the main text. We keep parameter $\gamma_0$ fixed, given overall agreement in the literature. Let us first discuss the qualitative effect of these parameters on the band structure and resulting DOS pattern. Decreasing $\gamma_3$ decreases the curvatures of bilayer bands at small momenta. This decreases the distance between the tip of feature (3), which is due to BLG-like band extrema, (see the main text) and the origin, see Fig.~\ref{g3}. Parameter $\Delta_2$ shifts most of the features on the electron doped side (and also Lifshits transitions at negative fillings, given roughly by (8) and (9)) away from the NP, see Fig.~\ref{D2}. Finally, Fig.~\ref{g4} illustrates the effect of changing $\gamma_4$. We observe that DOS is not very sensitive to  $\gamma_4$  which has the most pronounced effect on the positions of Lifshits points (8) and (9) on the hole-doped side.
	
	The above intuition suggests that parameters $\Delta_2$ and $\gamma_3$ has to be respectively increased and decreased compared to their values in Ref.~\cite{serbyn_new_2013}. We determine the values of $\Delta_2$ and $\gamma_3$ which give the closest agreement between our simulation and experimental data to be
	\begin{equation}\label{Eq:g3D2}
	\gamma_3=0.29~\text{eV},
	\quad
	\Delta_2 = 3.5\pm0.2~\text{meV},
	\end{equation}
	where we estimated error bar for $\Delta_2$ from the sensitivity of Landau levels plot. Due to very weak effect of $\gamma_3$ on LL crossing pattern, we could not quantify the associated error bars. However, Fig.~\ref{g3} suggests that changing $\gamma_3$ by $0.05$~eV visibly degrades agreement of our simulations with experimental data.

	Finally, Fig.~\ref{g4} shows the effect of changing $\gamma_4$. Increase in $\gamma_4$ brings Lifshits transitions on the hole doped side closer to each other. This removes the dip in the DOS that would be present otherwise between Lifshits transition at small values of $\gamma_4$, and which is not observed in the experiment. Since the experimental data does not allow for a very precise determination of Lifshits points, it is hard to estimate the error bar on our value  $\gamma_4=0.141$~eV. At the same time, we can estimate error bar for $\gamma_4$ using its effect on the position of the triplet crossing, following a procedure similar to that for $\gamma_5$, as:
	\begin{equation}\label{Eq:g4sm}
	\gamma_4=0.141 \pm 0.04~\text{eV}.
	\end{equation}

	Collecting together value ranges of tight-binding parameters in Eqs.~(\ref{Eq:g2sm})-(\ref{Eq:g4sm}) we arrive to the tight-binding parameter set
	\begin{subequations}
		\begin{eqnarray}\label{Eq:g2g5d}
		\gamma_1&=&0.380\pm 0.003~\text{eV},\\
		\gamma_2&=&-0.020\pm 0.005~\text{eV},\\
		\gamma_3&=&0.29~\text{eV},\\
		\gamma_4&=&0.141 \pm 0.04~\text{eV},\\
		\gamma_5&=&0.050\pm 0.005~\text{eV},\\
		\delta&=& 0.0355\pm0.0045~\text{eV},\\
		\Delta_2 &=& 3.5\pm0.2~\text{meV}.
		\end{eqnarray}
	\end{subequations}
	as the best set of parameters satisfying all constraints.
	listed in the last row of Table~\ref{TB}. Finally, we would like to point out that despite the overall agreement in positions of all features between experiment and our simulations, we were unable to obtain the correct \textit{magnitude} of DOS $\nu(\mu)$ between the two LPs at negative densities. The simulation values of DOS far exceed the experimentally observed values. We attribute this disagreement to possible interaction effects which may be enhanced due to the proximity of two Lifshits points.

	\section{C. Effect of interactions: Hartree Fock approximation}
	
	\subsection{C1. Symmetry broken states in emergent triplets}
	
	In this Section we describe the Hartree-Fock (HF) approximation for   completely filled Landau Levels (LL) originally proposed in Ref.~\cite{macdonald_influence_1984}. The essence of the method is a variational optimization of the energy over a trial set of wave functions (Slater determinants). In this work we largely follow approach of Ref.~\cite{zhang_hunds_2012}.  We aim to capture the interactions-induced splitting of emergent (nearly) three-fold degenerate Landau levels formed at large $\Delta_1$. In what follows we refer to such states as ``triplets'', where three-fold degeneracy originates from the set of three Dirac cones related to each other via $C_3$ rotation symmetry, see Fig.~2g in the main text. Hence, we restrict our set of variational states to an arbitrary superpositions of single-particle triplet wave functions.
	
	More specifically, we start with the set of six Landau level wave functions denoted as $\psi^{(ms)}_{\text{tri}}$, $m=1,2,3$. Index $s$ labels spin projection onto $z$-axis, so that $\psi^{(m\uparrow)}_{\text{tri}} =\psi^{(m)}_{\text{tri}}\otimes |\uparrow\rangle$ and $\psi^{(m,\downarrow)}_{\text{tri}} =\psi^{(m)}_{\text{tri}}\otimes |\downarrow\rangle$, with the wave function $\psi^{(m)}_{\text{tri}}$ obtained from exact diagonalization of Hamiltonian~(\ref{Hamiltonian-full}). Three states $\psi^{(m)}_{\text{tri}}$ with $m=1,2,3$ can be distinguished by their transformation under $C_3$ rotations which can be intuitively seen as a proxy of ``angular momentum''. Due to presence of discrete rotational symmetry, this ``angular momentum'' is defined modulo 3 and takes values $0$, $1$, and $2$, corresponding to phase of $0$, $2\pi/3$ and $4\pi/3$ acquired from rotation by angle of $2\pi/3$.
	
	The wave functions $\psi^{(m)}_{\text{tri}}$ are vectors in the basis of Landau level indices and sublattices. Note, that the valley indices are omitted since all 3 Landau level forming the triplet belong to the same valley. In addition, we introduce a LL index cut-off $\Lambda_{\text{max}}=12$  which allows to represent triplet vector norm of about $\sim0.9$, thus incorporating most of the tripltets weight.
	
	Projecting Hamiltonian on the manifold of 6 triplet states, we get the following expression for the projected Hamiltonian:
	\begin{multline}\label{HF1}
	\langle m, s|H|m', s' \rangle = E_0(m)\delta_{m,m'}\delta_{s,s'}-E_{ZM} \sigma^z_{ss'} \\
	+(U_H)^{ms}_{m's'}+ J^{ms}_{m's'}.
	\end{multline}
	In this Hamiltonian,  $E_0(m)$ represents the diagonal spin-degenerate single-particle Hamiltonian. The second term is the Zeemann energy which retains its standard form after projection onto the triplet states. The last two terms in Eq.~(\ref{HF1}) originated from the interactions and account for Hartree and exchange terms respectively. These terms can be obtained from the rotation of conventional  Hartree and exchange terms by the wave functions of triplet states, and they depend on the density matrix in the basis of sublattices ($\alpha,\alpha'$) and  Landau levels ($n,n'$), $\Delta^{\alpha' n' s'}_{\alpha n s}$. This density matrix can be straightforwardly obtained from the density matrix in the triplet basis, $\Delta^{m_is_i}_{m_ks_k}$ via change of basis:
	\begin{equation}\label{DM}
	\Delta^{\beta n' s'}_{\alpha n s} = \sum_{m_i,m_k,s_i,s_k} \Delta^{m_is_i}_{m_ks_k}\psi^{(m_is_i)}_{\alpha n s} \otimes \psi^{(m_ks_k)\dagger}_{\beta n' s'}.
	\end{equation}
	
	Using density matrix in the basis of Landau levels, $\Delta^{\beta n' s'}_{\alpha n s}$, we can write standard expressions for Hartree and exchange terms, following Ref.~\cite{zhang_hunds_2012}:
	\begin{eqnarray}\label{ABAHamiltonian}
	\langle \alpha n s |U_{H}|\beta n' s'\rangle &=& \frac{E_H}{2}\Delta_\text{mid} (2\delta_{B_2,\alpha}+2\delta_{A_2,\alpha}-1),
	\\
	\langle \alpha n s |U_{ex}|\beta n' s'\rangle &=& J^{\alpha\beta s s'}_{n,n_1,n_2,n'}\Delta^{\beta n_2 s'}_{\alpha n_1 s}.
	\end{eqnarray}
	where parameter $E_H$,
	\begin{equation}
	E_H =\frac{e^2 d}{2 l^2_B \kappa},
	\end{equation}
	characterises the scale of the Hartree energy. $\kappa$ is the effective screening constant, $l_B = \sqrt{{\hbar c}/({eB})}$ is the magnetic length and $d=0.335$~nm measures the distance between adjacent graphene layers. Density matrix projection $\Delta_\text{mid} = \sum_{n,s} (\Delta^{A_2 n s}_{A_2 n s}+\Delta^{B_2 n s}_{B_2 n s})$ corresponds to the electron density on the middle layer. The exchange integral is defined as:
	\begin{equation}
	J^{\alpha\beta s s'}_{n,n_1,n_2,n'} = \int \frac{d^2q}{(2\pi)^2} U_{\alpha \beta}(q) F_{n,n_1}(-q) F_{n_2,n'}(q)  \delta_{ss'}.
	\end{equation}
	The explicit form of the form factors $F_{nn'}(q)$ is listed in Ref.~\cite{macdonald_influence_1984}, and the interaction potential in the exchange integral is given by:
	\begin{equation}\label{Potential}
	U_{\alpha\beta}(q) =  \frac{2\pi e}{q \varepsilon(q)} T_{\alpha \beta}
	\end{equation}
	where $\varepsilon(q)$ is the dielectric function. $T_{\alpha \beta} = 1, \exp (-qd)$ or $\exp(-2qd)$ for $\alpha,\beta$ in the same, adjacent or  different outer layers.
	
	The projection of the exchange interaction matrix onto the triplet basis is given by:
	\begin{equation}
	J^{m_is_i}_{m_ks_k} = \sum \psi^{(m_ks_k)}_{\beta n' s'} \langle \alpha, n, s |U_{ex}|\beta, n', s'\rangle  \psi^{(m_is_i)\dagger}_{\alpha n s},
	\end{equation}
	where the summation is taken over repeated indices. The same procedure must be applied to the Hartree term to obtain $(U_H)^{m_is_i}_{m_ks_k}$.

	The self-consistent solution of HF equations is implemented  as follows. For instance, fixing filling at $N=1$, we start with the trial density matrix in the triplet basis, $\Delta_{m_is_i,m_ks_k} = (c_1,c_2,c_3)\times (c_1,c_2,c_3)^T |\uparrow \rangle \langle \uparrow|$, where $c_i$ are random normalized coefficients, $\sum_{i=1}^3 |c_i|^2=1$. Using this density matrix, we calculate the density matrix in LL basis and exchange integrals according to Eqs.~(\ref{DM})-(\ref{Potential}). Finally, by diagonalizing projected Hamiltonian in  Eq.~(\ref{HF1}) we calculate updated eigenstates $|n\rangle$ and produce a new density matrix $\Delta^{m_is_i}_{m_ks_k}$ by filling the lowest $N$ of them,
	\[
	\Delta^{m_is_i}_{m_ks_k} = \sum_{n=1}^N |n \rangle \langle n|.
	\]
	The above procedure is repeated until the eigenvalues converge.
	
	We apply the above self-consistent HF procedure to the case of filling $N=1$ of the triplet T2 (see Fig 3d in the main text). We use the constant dielectric function $\varepsilon(q)=6.6$ and $\kappa =\varepsilon$. The calculation yields the symmetry broken state as the one which has the lowest variational energy. This symmetry broken states consists of superposition of all three single-particle triplet wave functions $\psi^{(m)}_\text{tri}$. Since each of the single-particle triplet wave functions acquires a different phase under $C_3$ rotation, such superposition of single particle wave functions breaks rotational symmetry.
	
	Intuitively, one can easily undertand why the interactions favor the symmetry broken state at $N=1$. Each of the single-particle wave functions $\psi^{(m)}_\text{tri}$, $m=1,2,3$ lives on all three Dirac points (see Fig.~3e in the main text). In fact, in the limit of weak magnetic field (or large separation between emergent Dirac gulleys), these single particle wave-functions become the proper combination of wave-functions localized on each of the Dirac cones $\phi_i$ with an additional phase factors
	\begin{eqnarray}
	\psi^{(1)}_\text{tri} &=& \frac{1}{\sqrt 3} (\phi_1+\phi_2+\phi_3),\\
	\psi^{(2)}_\text{tri} &=& \frac{1}{\sqrt 3} (\phi_1+e^{2\pi i/3}\phi_2+e^{4\pi i/3}\phi_3),\\
	\psi^{(3)}_\text{tri} &=& \frac{1}{\sqrt 3} (\phi_1+e^{4\pi i/3}\phi_2+e^{2\pi i/3}\phi_3).
	\end{eqnarray}
	The $C_3$ rotations simply permutes $\phi_i$ between themselves. This results in the function $\psi^{(1)}_\text{tri}$ being invariant under rotation, and remaining two states $\psi^{(2,3)}_\text{tri} $ acquiring a phase factor $e^{\pm 2\pi i/3}$. Now, since support of wave functions $\phi_i$ and $\phi_j$ are weakly overlapping for $i\neq j$, exchanges favor the state where all weight of the wave function is located in one of the Dirac gulleys. In the basis of $\psi^{(m)}_\text{tri}$ such state corresponds to a coherent superposition of all three single-particle wave functions and it breaks $C_3$ rotation symmetry.
	
	\subsection{C2. Visualizing symmetry broken states}
	
	In order to visualize the form of the symmetry broken states in real space, we transform the LL wave functions into the maximally localized ``wave packet''. This is done via convolving the single particle LL wave function in the Landau gauge with the Gaussian envelope function,
	\[
	\Psi_{n} (x,y) = \int_{-\infty}^{\infty} C_X \exp(i X y / l_B^2) \psi_n\bigg(\frac{x-X}{l_B}\bigg) dX
	\]
	where $\psi_n$ is the $n$-th eigenstate of the Hamiltonian. In order to get the maximally localized wave packet in both directions, we choose $C_X = (2\pi l_B^2)^{-\frac{1}{2}} \exp(-X^2/2l_B^2)$. We calculate the integral using explicit expression for $\psi_n$,
	\[
	\psi_n(x) = \frac{1}{\pi^{\frac{1}{4}} \sqrt{2^n n! l_B}} \exp(- x^2 /2) H_n(x),
	\]
    where $H_n(x)$ is the $n$-th Hermite polynomial. This gives the following wave function describing LL ``wave packet'' centered at the origin:
	\begin{multline}
	\Psi_{n} (x,y) = \frac{1}{\sqrt{n!}} \bigg(\frac{x-iy}{\sqrt{2}l_B}\bigg)^n \\
	\times \exp \bigg(-\frac{x^2+y^2}{4 l_B^2}\bigg) \exp \bigg(i \frac{xy}{2 l_B^2}\bigg).
	\end{multline}
	
	We numerically simulate the probability distribution for the triplet eigenstates $\psi^{(m)}_\text{tri}$, $m=1,2,3$ at $B=1.25$~T and compare them with the momentum band structure. More specifically, we plot probability density $p(x,y)$ for the wave function in the basis of LL and sublattices, $\psi^{\alpha n}$, is calculated as
	\begin{equation}
	p(x,y)
	=
	\sum_{\alpha=1}^6\left|
	\sum_{n=1}^{\Lambda_\text{max}} \psi^{\alpha n} \Psi_{n} (x,y)
	\right|^2,
	\end{equation}
	where the inner sum goes over LL and outer sum sums probability density for each of the sublattices. The probability density calculated for the single-particle triplet wave functions is shown in Fig.~3e in the main text. Indeed, as expected we observe that maximas of $p(x,y)$ are centered around their spatial semiclassical trajectories which coincide with the position of Dirac gullies in momentum space after $\pi/2$ rotation.
	
	Figure~3f in the main text shows $p(x,y)$ for the self-consistent eigenstate at $B=1.25$~T and $\Delta_1=0.08$. From this plot it is clear that the HF state breaks $C_3$ symmetry as it is strongly localized in a single Dirac gully.	
\clearpage

\newpage

\subsection*{Supplementary figures}
\begin{figure}[h]
\includegraphics[width=\linewidth]{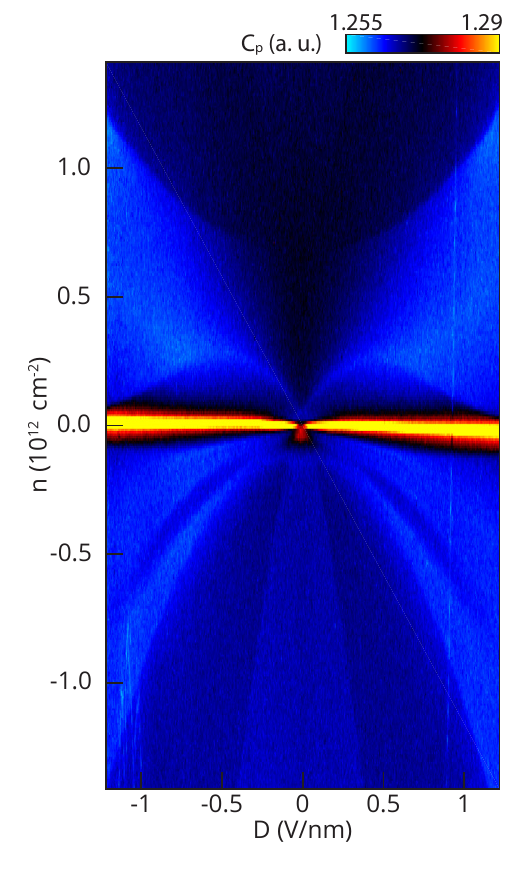}
\caption{\label{b0_hc} Penetration field capacitance $C_p$ at $B=0$~T and $T\approx 50$~mK as a function of $n$ and $D$ .}
\end{figure}
\begin{figure}[h]
\includegraphics[width=\linewidth]{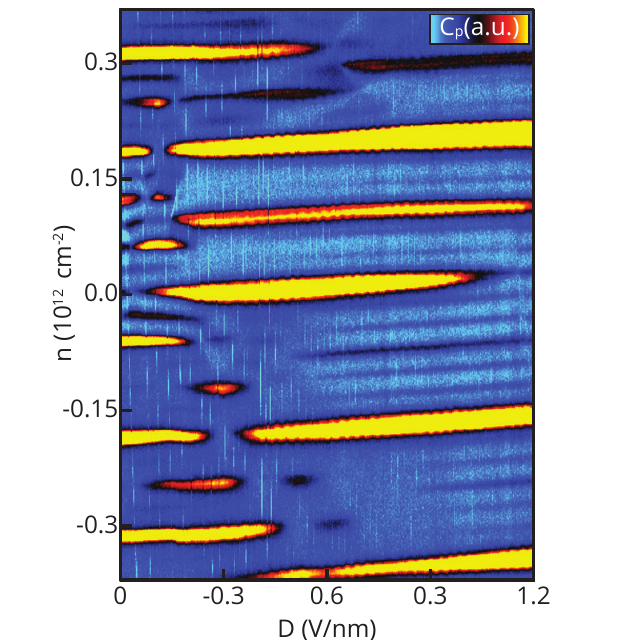}
\caption{\label{tripletzoom} Penetration field capacitance $C_p$ at $B=1.25$~T inisde the dashed region of Fig3a of main text showing symmetry broken quantum Hall states in the 'gully' regime}
\end{figure}

\end{document}